\documentclass[aps,pra,reprint,superscriptaddress,showpacs]{revtex4-2}
\usepackage{amsmath,amsfonts}
\usepackage[colorlinks,linkcolor=blue, urlcolor=blue, anchorcolor=blue, citecolor=blue]{hyperref}
\usepackage[T1]{fontenc}
\usepackage{mathptmx}
\usepackage{graphicx}
\usepackage{amsmath}
\usepackage{amssymb}
\usepackage{setspace}
\DeclareMathAlphabet{\mathcal}{OMS}{cmsy}{m}{n}  

\begin{document}
	
	
	\title{Quantum phase transition in XXZ central spin model}
	
	
	\author{Lei Shao}
	\author{Rui Zhang}
	\affiliation{Zhejiang Institute of Modern Physics, Department of Physics, Zhejiang University, Hangzhou 310027, China}
	\author{Wangjun Lu}
	\affiliation{Department of Maths and Physics, Hunan Institute of Engineering, Xiangtan 411104, China}
	\author{Zhucheng Zhang}
	\author{Xiaoguang Wang}
	\email{xgwang1208@zju.edu.cn}
	\affiliation{Zhejiang Institute of Modern Physics, Department of Physics, Zhejiang University, Hangzhou 310027, China}


	\begin{abstract}
		We investigate the quantum phase transition (QPT) in the XXZ central spin model, which can be described as a spin-$\frac{1}{2}$ particle coupled to $N$ bath spins. In general, the QPT is supposed to occur only in the thermodynamical limit. In contrast, we present that the central spin model exhibits a normal-to-superradiant phase transition in the limit where the ratio of the transition frequency of the central spin to that of the bath spins and the number of the bath spins tend to infinity. We give the low-energy effective Hamiltonian analytically in the normal phase and the superradiant phase, and we find that the  longitudinal interaction $\Delta$ can significantly influence the excitation number and the coherence of the ground state. These two quantities are remarkably enhanced for the negative longitudinal interaction while suppressed for the positive longitudinal interaction. We also use the quantum Fisher information (QFI) to characterize the QPT and illustrate a measurement scheme that can be applied in practice. This work builds a novel connection between the qubit-spin systems and the qubit-field systems, which provides a possibility for the realization of criticality-enhanced quantum sensing in central spin systems.
	\end{abstract}
	
	
	\maketitle
	
	\section{Introduction}
	Quantum phase transition of many-body systems plays an important role in our understanding of physics~\cite{sachdev_2011}. While Landau’s symmetry-breaking theory gains great success in describing thermal phase transitions, quantum phase transitions, which is due to quantum fluctuations when temperature goes to zero beyond the symmetry-breaking paradigm, attracts a lot of interest in condensed matter  physics~\cite{PhysRevLett.95.105701,PhysRevLett.95.245701,PhysRevLett.110.135704,PhysRevB.58.R14741,Greiner2002,Vojta_2003,PhysRevLett.95.105701,RevModPhys.69.315,PhysRevB.75.134302} and quantum optics~\cite{PhysRev.93.99,HEPP1973360,Greentree2006,GROSS1982301,PhysRevA.7.831,PhysRevLett.119.220601,PhysRevA.101.033827,Cai2021,PhysRevLett.127.063602}. As a well-known model in quantum optics, the Dicke model~\cite{PhysRev.93.99} describes $N$ two-level atoms coupled to a single-mode cavity, and a quantum phase transition from the normal phase to the superradiant phase will occur in the thermodynamic limit $N\rightarrow\infty$. Recently, Hwang \emph{et al}.~\cite{PhysRevLett.115.180404,PhysRevLett.117.123602} presented that QPT can occur under the situation of a two-level atom coupled to a single-mode cavity, and they indicated that the ratio $\eta$ of the atomic transition frequency to the cavity field frequency plays the same role in the quantum Rabi model (QRM) and the Jaynes-Cummings model (JC) as the number of atoms in the Dicke model and the Tavis-Cummings model~\cite{PhysRev.170.379}.
	They also showed that the two-site JC lattice undergoes a Mott-insulating-superfluid QPT in the limit $\eta\rightarrow\infty$~\cite{PhysRevLett.117.123602}.

	Actually, there exist a similarity between the light-matter interaction in optical systems and the hyperfine interaction in spin systems. Like the phenomenon of  superradiance in quantum optics mentioned earlier, the superradiant effect can occur in the nuclear spin environment~\cite{PhysRevLett.104.143601,PhysRevB.86.085322,PhysRevB.91.245306,PhysRevB.103.155301}. Kessler \emph{et al}.~\cite{PhysRevLett.104.143601} showed that the superradiant effect can be realized in the systems that nuclear spin ensemble surrounding a quantum dot or an nitrogen-vacancy (NV) center, and  some suitable optical pumping conditions are given. In Ref.~\cite{PhysRevA.87.052323}, Dooley \emph{et al}. exploited the spin coherent state as the initial state to  discuss the collapse and revival phenomena in qubit-big spin model and revealed the similarities of the Hamiltonian between the qubit-spin systems and the qubit-field systems. In Ref.~\cite{PhysRevB.99.174308}, He \emph{et al}. gave the exact quantum dynamics of the XXZ central spin model and the analytic expression of quantum collapse and revival is also obtained. Moreover, they used the Holstein-Primakoff transformation to build a mapping between the  central spin model and the JC model. Furthermore, the connection between the spin-$s$ central spin model and Tavis-Cummings model is discussed in Ref.~\cite{Nepomechie_2018}.
	
    \setlength{\parskip}{1em} 
    
	However, the superradiant QPT has not been widely discussed in anisotropic central spin systems. Inspired by the work mentioned above, we follow the thoughts in Refs.~\cite{PhysRevLett.115.180404,PhysRevLett.117.123602,PhysRevB.99.174308} and analytically analyze the superradiant quantum phase transition in the XXZ central spin model. 
	We first give the exact energy spectrum of the XXZ central spin model, and the asymptotic behavior between the central spin model and the JC model can be clearly observed in the case of a large number of the bath spins.  The strength of longitudinal interaction $\Delta$ in the XXZ central spin model is  different from that of the transverse interaction $A$, which can significantly influence the critical point of the phase transition. Thus, it is necessary to discuss the two following cases: (\romannumeral1) $\Delta=0$, (\romannumeral2) $\left|\Delta\right|<\omega$, where $\omega$ is frequency of the bath spins. It shows that the XXZ central spin model has a similar critical point with the JC model under the condition of $\Delta=0$. However, for $
	\left|\Delta\right|<\omega$, the critical point is  different from the one before and we give an analytical solution for this by means of the theory of low-energy effective Hamiltonian.

	This article is organized as follows. In Sec.~\ref{sec:level2}, we give the exact energy spectrum of the XXZ central spin model and  obtain the analytic expression of excitation number via the mean-field approximation. In Sec.~\ref{sec:level3}, we present the derivation of the low-energy effective Hamiltonian and exploit it to analyze the critical point and the ground state energy. In Sec.~\ref{sec:level4}, we discuss the influence of the longitudinal interaction $\Delta$ on the excitation number and the coherence of the ground state. In Sec.~\ref{sec:level5}, we make use of the QFI to characterize the QPT of the XXZ central spin model, and give a measurement scheme. Finally, we give a conclusion in Sec.\ref{sec:level6}.

	\section{\label{sec:level2}MODEL}
	The central spin model can be described as a single spin-$\frac{1}{2}$ particle coupled to $N$ spin-$\frac{1}{2}$ particles (bath spins), which is also called  qubit-big spin model in Ref.~\cite{PhysRevA.87.052323}. For different strengths of the longitudinal and the transverse interactions, the Hamiltonian of this model can be written as~\cite{PhysRevB.99.174308} (we set $\hbar=1$)
    \begin{equation}\label{E1}
	\begin{split}
	H=&\frac{\omega_{0}}{2}\sigma^{(0)}_{z}+\frac{\omega}{2} \sum_{k=1}^{N}\sigma_{z}^{(k)}+
	\sum_{k=1}^{N}\frac{\Delta_{k}}{2} \sigma^{\left(k\right)}_{z}\sigma^{\left(0\right)}_{z}\\
	&+\sum^{N}_{k=1}\frac{A_{k}}{2}\left(\sigma^{(k)}_{x}\sigma^{\left(0\right)}_{x}+\sigma^{\left(k\right)}_{y}\sigma^{\left(0\right)}_{y}\right),
	\end{split}
	\end{equation} 
	where $\omega_{0}$ and $\omega$ are respectively the transition frequency of the central spin and bath spins, $A_{k}$ is the strength of transverse interaction, and $\Delta_{k}$ is the longitudinal interaction. $\sigma^{0}_{i}$ denotes the Pauli operator of the central spin and $\sigma^{(k)}_{i}$ ($i=x,y,z$) denotes the Pauli operator of the bath spins. The central spin model is widely applied to solve the problem in quantum dots~\cite{PhysRevLett.88.186802,PhysRevB.76.045312,PhysRevLett.88.186802,PhysRevB.82.161308,PhysRevB.76.014304} and NV centers~\cite{PhysRevLett.118.150504}, and the bath spins can be regarded as a big spin or a quantum reference frame~\cite{PhysRevA.82.032320,PhysRevA.85.022333,_afr_nek_2015}, which can be promising in quantum metrology~\cite{PhysRevLett.96.010401,Giovannetti2011,PhysRevLett.121.020402,PhysRevA.78.042105,PhysRevLett.121.020402,PhysRevE.76.022101,PhysRevA.78.042106,PhysRevE.93.052118}. For simplicity, we consider the case that the central spin is uniformly coupled to bath spins ($A_{k}=A$, $\Delta_{k}=\Delta$), and the Hamiltonian becomes 
	\begin{equation}\label{E2}
	H=\omega_{0} S_{z}+\omega J_{z}+A\left(J_{+}S_{-}+J_{-}S_{+}\right)+2\Delta J_{z}S_{z},
	\end{equation}
	where $S_{z}=\frac{1}{2}\sigma_{z}^{(0)}$, $J_{z}=\sum_{k=1}^{N}\sigma_{z}^{(k)}$, and $N$ bath spins can be equivalent to a big spin with spin $j$ ($j=N/2$). From Eq.~(\ref{E1}), we see that the Hamiltonian of the XXZ central spin model exhibits a $U(1)$ symmetry. 
	Now we introduce the Dicke states $\left|N/2,m\right>=\left|j,m\right>$ ($m\in\left[-j,j\right]$) as the eigenstates of $J_{z}$ and the $\left|\uparrow\right>$ ($\left|\downarrow\right>$) as the eigenstate of $\sigma_{z}^{(0)}$. 
	
	In this paper, we denote $\left|\uparrow(\downarrow)\right>\otimes\left|j,n-j\right>$ as 
	$\left|\uparrow(\downarrow),n\right>$ ($n\in[0,2j]$), and $n$ can be viewed as the excitation number of bath spins. Note that the $\left|\uparrow,n-1\right>$ and $\left|\downarrow,n\right>$ are analogous to the bare states \cite{gerry_knight_2004} of the JC model. Due to the $U(1)$ symmetry of the Hamiltonian in Eq.~(\ref{E2}), it is easy to obtain the energy eigenvalues
	\begin{equation} \label{E3}
	E_{\pm}=\frac{1}{2}\left[\!\!\left(2m+1\right)\omega-\Delta\pm\sqrt{\!\left(\!\left(2m+1\right)\Delta\!-\omega+\omega_{0}\right)^{2}\!\!+\!4A^{2}k_{n}}\right]
	\end{equation}
	where $m=n-1-j$, $k_{n}=2j-n+1$. More detailed derivations are presented in Appendix \ref{Appendix:A}. 
	
	Up to now we have not discussed the value range of the parameter $\Delta$. Note that $\left|\downarrow,0\right>$ is the ground state of Eq.~(\ref{E2}) for $\left|\Delta\right|<\omega$, and its eigenvalue is $E_{\downarrow,0}=-\frac{\omega_{0}}{2}-\left(\omega-\Delta\right)j$. In the subsequent sections, we will prove $\omega<\left|\Delta\right|$ is a necessary condition for the superradiance QPT in the XXZ central spin model and we will also discuss the influence of different $\Delta$ on the XXZ central spin model.
	Now we consider the $E_{-}$ in the limit $\eta=\omega_{0}/\omega\rightarrow\infty$ and $g=A\sqrt{2j}/\sqrt{\omega_{0}\omega}=\lambda/\sqrt{\omega_{0}\omega}$. In order to make $g$ satisfies that $g\sim \mathcal{O}(1)$,  we need to ensure that $\lambda/\omega\sim \sqrt{\eta}$. 
	For $\Delta=0$ and $\eta\gg1$, $E_{-}$ in Eq.~(\ref{E3}) can be expanded into 
	\begin{equation}\label{E4}
	E_{-}=-\frac{\omega_{0}}{2}-\omega j+\left(1-g^{2}+g^{2}\frac{n-1}{2 j}\right)n\omega.
	\end{equation}
	In the limit $ N\rightarrow\infty$, the nonlinear term in Eq.~(\ref{E4}) can be negligible and  we obtain $ E_{-}=\left(1-g^2\right)n\omega+E_{\downarrow,0}$, which has a similar harmonic spectrum presented in Ref.~\cite{PhysRevLett.117.123602}. For $g<1$, $E_{-}$ is minimum at $n=0$ and the ground state energy  in the normal phase is $-\omega_{0}/2-\omega j$. For $g=1$, there exist a degeneracy between $ \left|\psi_{-}\left(n\right)\right\rangle $ (Eq.~(\ref{A4})) and $\left|\downarrow,0\right>$ and the normal-to-superradiant phase transition occurs at this critical point. It is clear to see that for $g>1$ the ground state is instable and its energy can decrease infinitely as the excitation number increases, and the bath spins are macroscopically excited just like the behavior of the cavity field in the JC model~\cite{PhysRevLett.117.123602}.
	
	Now we calculate the excitation number of the ground state. For $N\gg n$ and $ \eta\gg1$, $E_{-}$ in  Eq.~(\ref{E3}) can be written as 
	\begin{equation}\label{E5}
	E_{-}=-\frac{\omega_{0}}{2}\sqrt{1+4g^{2}n\eta^{-1}}+\omega_{0}\left(n-j-1\right)\eta^{-1},
	\end{equation} 
	and utilize $\left(\partial{E_{-}}/\partial{n}\right)/\omega_{0}=0$, we find that  the excitation number of the ground state  is $ n_{\rm{g}}=0 $ for $ g<1 $. For $ g>1 $, the excitation number of the ground state is given by
	\begin{equation}\label{E6}
	n_{\rm{g}}=\frac{\eta}{4}\left(g^2-g^{-2}\right),
	\end{equation}
	which is consistent with the result in Ref.~\cite{PhysRevLett.117.123602}.
	
	It is hard to acquire an analytical expression for $\left|\Delta\right|<\omega$ from Eq.~(\ref{E3}), thus we use the mean-field approximation to get the mean-field energy, which is given by
	\begin{equation}\label{E7}
	E_{-}=\omega\left(n-j\right)-\frac{\overline{\omega}_{0}(n)}{2},
	\end{equation}
	where $ \overline{\omega}_{0}(n)=\sqrt{4\lambda^{2}n+4n^{2}\Delta^{2}+4n\Delta\widetilde{\omega}_{0}+\widetilde{\omega}_{0}^{2}} $, $\widetilde{\omega}_{0}=\omega_{0}-N\Delta$, and $\widetilde{\omega}=\omega-\Delta$.

    In the normal phase, the excitation number of the ground state  is still $ n_{\rm{g}}=0$  . But for the superradiant phase, the excitation number is given by
	\begin{equation}\label{E8}
	n_{\rm{g}}=-\frac{\lambda^{2}+\text{\ensuremath{\Delta\widetilde{\omega}_{0}}}}{2\Delta^{2}}+\frac{\lambda\omega}{2\Delta^{2}}\sqrt{\frac{\lambda^{2}+2\Delta\widetilde{\omega}_{0}}{\omega^{2}-\Delta^{2}}},
	\end{equation}
	and the ground state energy is 
	\begin{equation}\label{E9}
		E_{g}^{\mathrm{MF}}=\left(n_{\mathrm g}-j\right)\omega-\frac{1}{2}\overline{\omega}_{0}(n_{\rm g})
	\end{equation}
	The detailed derivation is presented in Appendix \ref{Appendix:A}. This result is completely different from the previous situation since the existence of the nonlinear coupling term~\cite{Eckle2017}.

	\section{\label{sec:level3}LOW-ENERGY EFFECTIVE HAMILTONIAN}
	
	In order to further understand the QPT in the XXZ central spin model, in this section we give the low-energy effective Hamiltonian both in the normal phase and the superradiant phase. Note that Eq.~(\ref{E2}) can be mapped to the Hamiltonian of the JC model when $\Delta=0$, which has been discussed in detail in Ref.~\cite{PhysRevLett.117.123602}, thus we focus on $\left|\Delta\right|<\omega$ in this paper.
	
	For the normal phase, we apply a Holstein-Primakoff transformation and a Schrieffer-Wolff transformation $e^{S}$ with the anti-Hermitian operator $S=\lambda\left(a^{\dagger} S_{-}-aS_{+}\right)/\widetilde{\omega}_{0} $~\cite{PhysRevLett.115.180404,PhysRevLett.117.123602} to Eq.~(\ref{E2}), and we obtain
	the low-energy effective Hamiltonian which is
	\begin{equation}\label{E10}
	\widetilde{H}_{\rm{np}}
	=-\frac{\widetilde{\omega}_{0}}{2}-\omega j+\widetilde{\omega}\left(1-\tilde{g}^{2}\right)a^{\dagger}a,
	\end{equation}
	where $a$ ($a^{\dagger}$) is the bosonic annihilation (creation) operator, and $\tilde{g}=\lambda/\sqrt{\widetilde{\omega}\widetilde{\omega}_{0}}  $. The detailed derivation of $\widetilde{H}_{\rm{np}}$ is presented in Appendix \ref{Appendix:B}. Equation (\ref{E9}) shows a similar structure with Eq.~(\ref{E4}) in the limit $ N\rightarrow\infty $. 
	
	Now we prove that $ \left|\Delta\right|<\omega $ is a necessary condition to acquire the QPT similar to the JC model. First of all, the comparison between Eq.~(\ref{E4}) and Eq.~(\ref{E9}) shows that $ \widetilde{\omega} $ should satisfy that $ \widetilde{\omega}>0 $, thus we have $ \Delta<\omega$.
	Secondly, $n_{\rm{g}}$ only makes sense if 
	\begin{equation}\label{E11}
	\frac{\lambda^{2}+2\Delta\widetilde{\omega}_{0}}{\omega^{2}-\Delta^{2}}>0,
	\end{equation}
	and it is easy to verify that Eq.~(\ref{E10}) can be always satisfied if $ \Delta>-\omega $ in the range of $ \tilde{g}>1 $. Finally, we get the necessary condition for the superradiance QPT in the XXZ central spin model is $\left|\Delta\right|<\omega$. 
	 
	 From Eq.~(\ref{E9})  we see that the new critical point for $\left|\Delta\right|<\omega$ is $\lambda_{c}=\sqrt{\widetilde{\omega}_{0}\widetilde{\omega}}$.
	For $\tilde{g}<1$, the ground state is $\left|\psi_{\rm{g}}^{\rm {np}}\right>=e^{S}\left|\downarrow,0\right>$ with energy $E_{\rm{np}}= -\widetilde{\omega}_{0}/2-\omega j $. Note that if $ \Delta<\omega $ is not satisfied, then the ground state is $\left|\uparrow,0\right>$, which is not we expect. . However, the XXZ central spin model exhibits the instability when $\tilde{g}>1 $, and to solve this problem, we use the method proposed in Refs.~\cite{PhysRevLett.115.180404,PhysRevLett.117.123602} to get the low-energy effective Hamiltonian in the superradiant phase.
	\begin{figure}[tbp]
		\includegraphics[width=88mm,height=116mm]{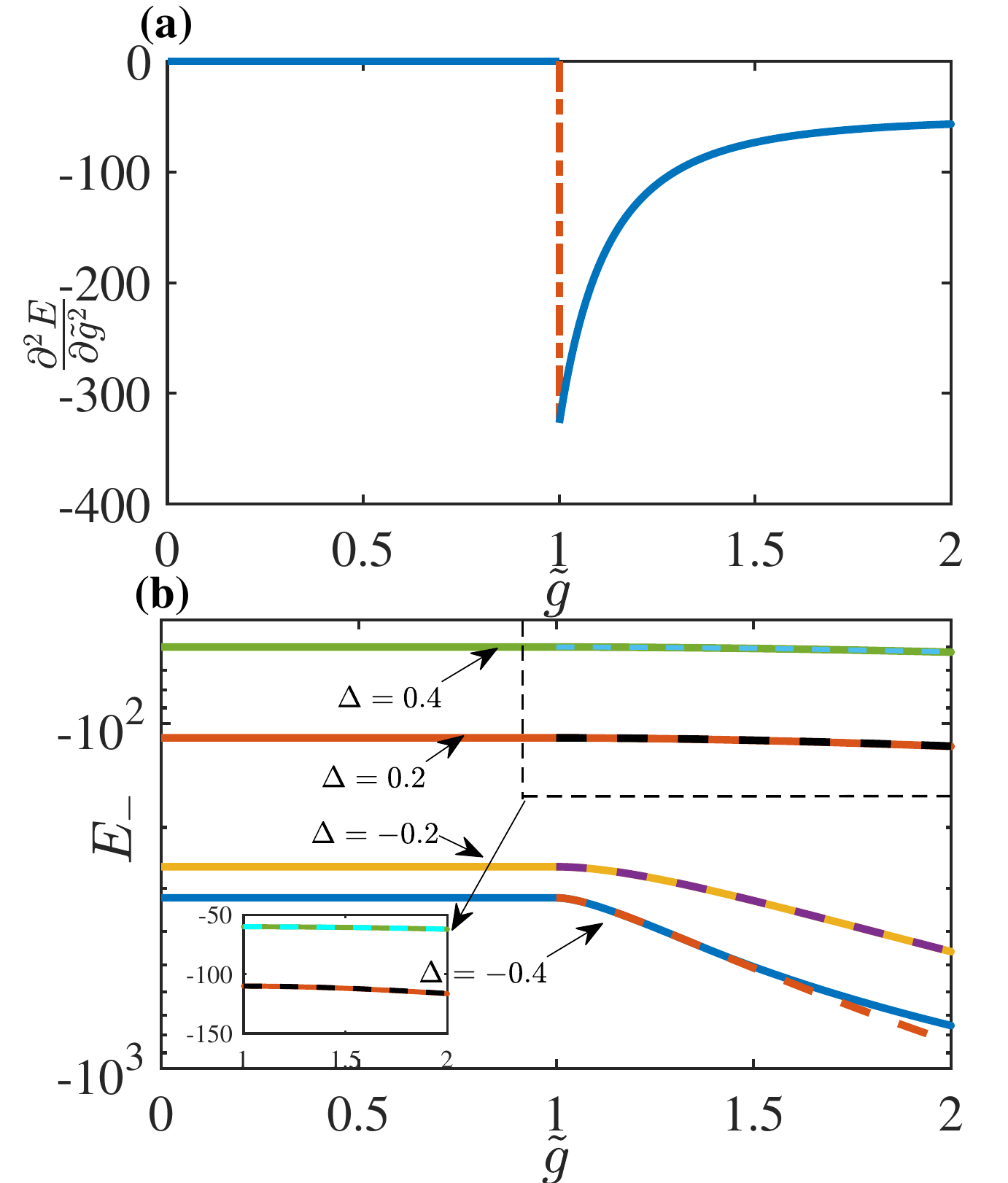}
		\caption{\label{fig:1} (a) Second derivative of the ground state energy with respect to $ \tilde{g} $. (b) Ground state energy of the XXZ central spin model for different longitudinal interactions $ \Delta $. Here we set $ \omega_{0}=100 $, $\omega=0.5 $. The solid lines are the result of  the numerical simulation and the dotted lines are that of the analytical expression in Eq.~(\ref{E9}). The top two curves (light blue line and black line ) fit better than the bottom two (purple line and red line).
		}
	\end{figure}
	
	Unlike the previous approach, we need to apply a displacement operator in addition to the two transformations mentioned before. Finally we get the low-energy effective Hamiltonian is 
	\begin{equation}\label{E12}
	\overline{H}'_{\rm {sp}}=\omega\alpha ^{2}-\frac{\overline{\omega}_{0}}{2}+\kappa_{0} x^{2}-\omega j,
	\end{equation}
	where 
	\begin{equation}\label{E13}
	\kappa_{0}=\frac{\omega}{4}-\frac{2\alpha ^{2}\Delta^{2}+\Delta\widetilde{\omega}_{0}}{4\overline{\omega}_{0}},
	\end{equation}
	$x=a^{\dagger}+a$, $ \alpha^2=n_{\rm g} $ and $\overline{\omega}_{0}=\sqrt{\frac{\lambda^{4}+2\Delta\lambda^{2}\widetilde{\omega}_{0}}{\omega^{2}-\Delta^{2}}}$. The detailed derivation is presented in Appendix \ref{Appendix:B}. The ground state of 
	$\widetilde{H}_{\rm{sp}}^{''}$ is $ \left|\psi_{g}^{\rm{sp}}\right> =\mathcal{D}\left(
	\alpha\right)e^{\overline{S}}\mathcal{S}\left(r\right)\left|\tilde{\downarrow},0\right>$, where $\mathcal{ D}\left(\alpha\right)=\exp\left[\alpha\left(a^{\dagger}-a\right)\right]$ and $\mathcal{S}\left(r\right)=\lim_{r\rightarrow\infty}\exp\left[\frac{r}{2}\left(a^2-a^{\dagger2}\right)\right]$. Its ground state energy is approximately equal to the mean-field energy in Eq.~(\ref{E9}). In Fig.~\ref{fig:1}(a), we show that the second derivative of the ground state energy is discontinuous at the critical point, which means the QPT occurs in the XXZ central spin model is a second-order phase transition like the JC model~\cite{PhysRevLett.117.123602}. In Fig.~\ref{fig:1}(b), we present that the agreement between the analytical expression and the numerical simulation of the ground state energy. It shows that  Eq.~(\ref{E9}) agrees well with the numerical result when $ \left|\Delta\right| $ is small relative to $ \omega $, however, as $ \left|\Delta\right| $ gets closer to $ \omega $, the agreement becomes worse. The main reason for this problem is that when $ \left|\Delta\right| $ is close to $ \omega $, a larger $ N $ is required for a better numerical simulation, which places a high demand on the memory of the computer.

	\section{\label{sec:level4}INFLUENCE OF LONGITUDINAL INTERACTION}\
	In this section, we will discuss the influence of the longitudinal interaction $\Delta$ on the excitation number $n_{\rm g}$. The excitation number can be regarded as an order parameter since it keeps zero in the normal phase, and becomes nonzero in the superradiant phase. For $ \Delta=0 $, $ n_{\rm g} $ can be described by Eq.~(\ref{E8}), however, it is different for $ -\omega<\Delta<0 $ and $ 0<\Delta<\omega $, thus we must discuss it separately. 
	\begin{figure}[tpb]
		
		\includegraphics[width=90mm,height=106mm]{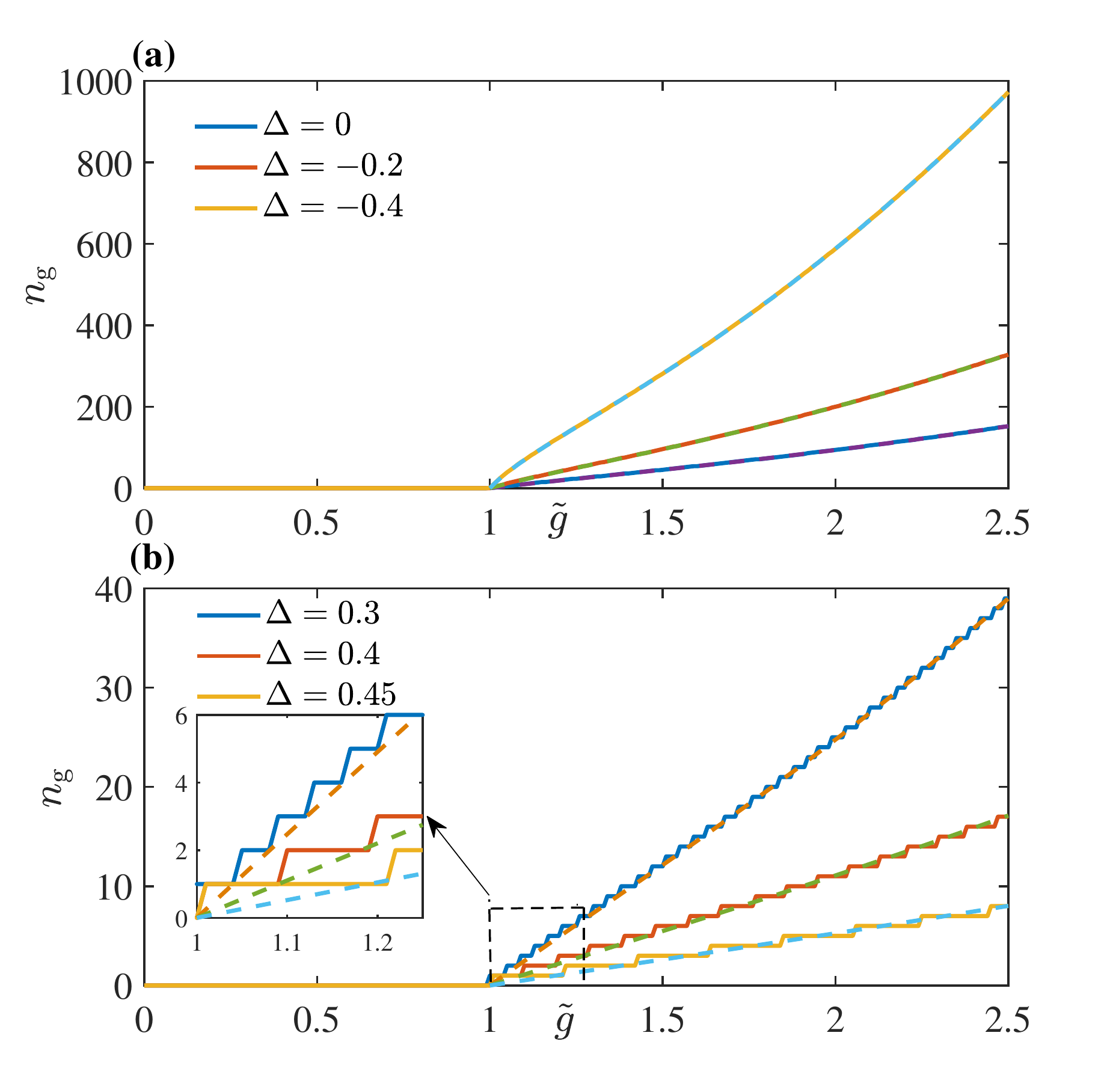}
		
		\caption{\label{fig:2}  Variations of the excitation number $n_{\rm g}$ with respect to $\tilde{g}$. The solid lines are the numerical simulation and the dashed lines are the analytical expression given in Eq.~(\ref{E8}). Here we set $ \omega=0.5 $ and $ \eta=100 $. In (a), the excitation numbers are significantly enhanced in the range of $ -\omega<\Delta<0 $. In (b), the excitation numbers are suppressed in the range of $0<\Delta<\omega $, and the curves become discontinued when $ \Delta $ gets close to $\omega  $.   }
	\end{figure}
	In Fig.~\ref{fig:2}(a), the analytical results given by Eq.~(\ref{E8}) agree well with the numerical results for different negative longitudinal interaction. Moreover, we find a significant increase in the number of excitations with the increase of the absolute value of the longitudinal interaction $\left|\Delta\right| $ in the range of $ -\omega<\Delta<0 $. We can explain this phenomenon in terms  of the analytical expression Eq.~(\ref{E8}), which can be rewritten as 

	\begin{equation}\label{E14}
	n_{\rm g}=\!\frac{\lambda^{2}\left(\lambda^{2}+2\Delta\widetilde{\omega}_{0}\right)-\left(\omega^{2}\!-\!\Delta^{2}\right)\widetilde{\omega}_{0}^{2}}{2\!\left(\omega^{2}\!-\!\Delta^{2}\right)\!\left(\lambda^{2}\!+\!\Delta\widetilde{\omega}_{0}\right)\!+\!2\lambda\omega\sqrt{\left(\omega^{2}\!-\!\Delta^{2}\right)\!\left(\lambda^{2}+2\Delta\widetilde{\omega}_{0}\right)}},
	\end{equation}
	and we can find that Eq.~(\ref{E14}) becomes Eq.~(\ref{E6}) under the condition of $ \Delta=0 $.  It is easy to see that the numerator remains finite when $ \Delta\simeq-\omega $, while the denominator tends to infinity. Therefore, compared with the situation of $ \Delta=0 $, the bath spins will be excited more quickly as $ \left|\Delta\right| $ approaches to $\omega $.  
	
	But for $ 0<\Delta<\omega$, the corresponding results are different. Figure~\ref{fig:2}(b) shows that the excitation number decreases as $ \Delta $ increases, and when $ \Delta $ is close enough to $ \omega $, the variation of the excitation number $ n_{\rm g} $ with respect to $ \tilde{g} $ becomes discontinuous (shape of a stair). Similarly, we seek an explanation from Eq.~(\ref{E8}).
	In the limit $\widetilde{\omega}=\omega-\Delta\rightarrow 0$, Eq.~(\ref{E8})  becomes 
	\begin{equation}\label{E15}
	n_{\rm g}=\frac{\widetilde{\omega}_{0}}{2\Delta^{2}}\left(\widetilde{g}\omega-\Delta\right).
	\end{equation}
	In Eq.~(\ref{E15}), for fixed $ \tilde{g} $, the numerator decreases ($ \tilde{g}>1$ and $\tilde{g}\omega>\Delta$) and the denominator increases with $ \Delta $ increases, thus $ n_{\rm g} $ decreases. Besides, the excitation number $n_{g}$ is insensitive  to $ \tilde{g} $ when $ \Delta$ is close to $ \omega $, therefore $ n_{\rm g} $ will vary discretely (shown in Fig.~\ref{fig:2}(b)). In summary, we see that the macroscopic excitations of bath spins are significantly enhanced in the case of the negative longitudinal interaction ($ -\omega<\Delta<0 $), but for the situation of the positive longitudinal interaction ($ 0<\Delta<\omega $), the excitations are suppressed.
		\begin{figure}[tbp]
		\includegraphics[width=90mm,height=66mm]{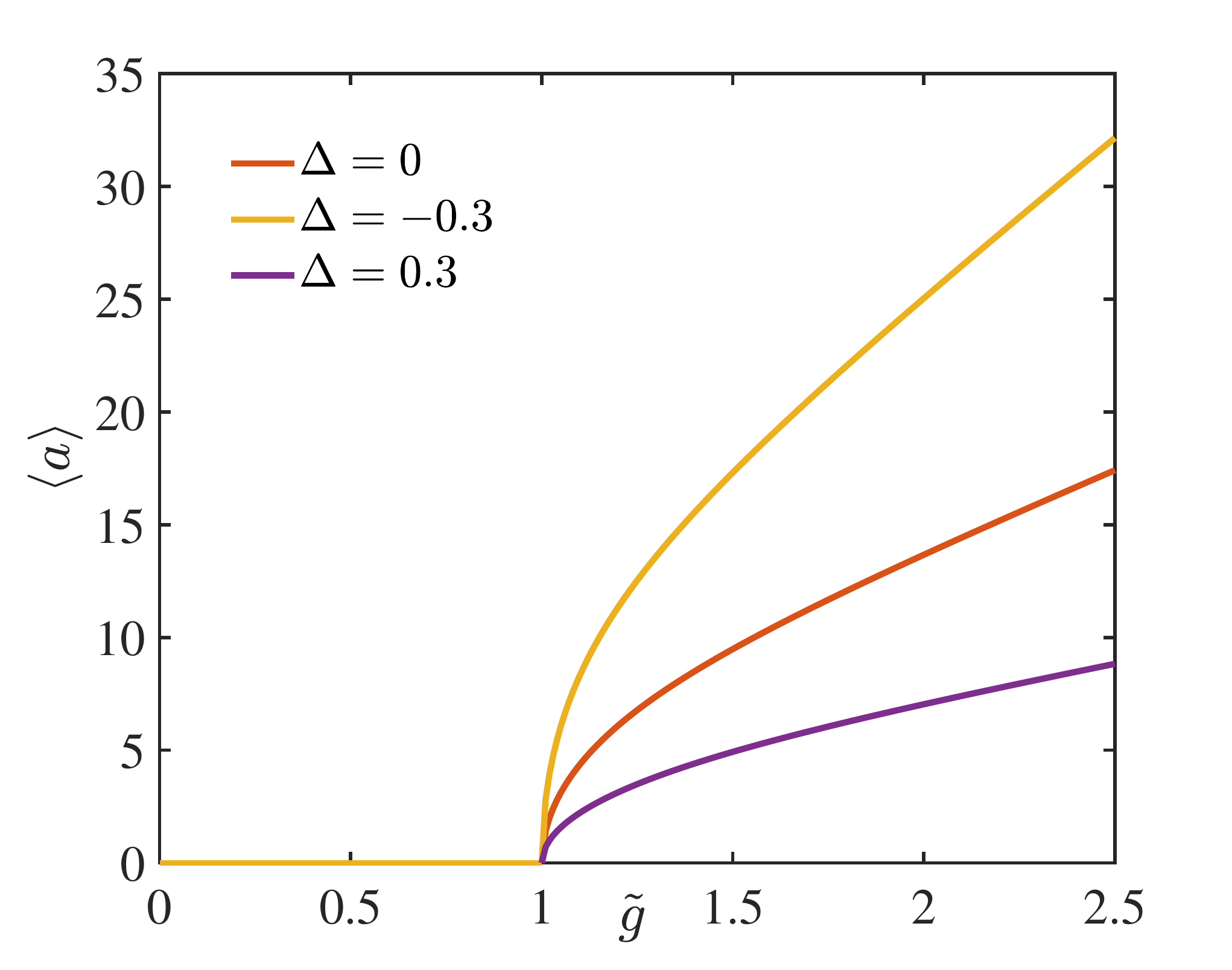}
		\caption{\label{fig:3} Coherence of the ground state 
			varies with $\tilde{g}$. For $0<\Delta<\omega$, the coherence  increases significantly, whereas for $-\omega<\Delta<0 $ the coherence decreases appreciably.    		}
	\end{figure}
	Moreover, for the superradiant phase, the ground state coherence of bath spins in the limit $ \eta\rightarrow\infty $ and $ N\rightarrow\infty $ is given by
	\begin{equation}\label{E16}
	\left<a\right>=\left<\psi_{g}^{\rm{sp}}\right| a\left|\psi_{g}^{\rm{sp}}\right> 
	=\sqrt{n_{\rm g}}.
	\end{equation}
	Similar to the excitation number of the ground state $ n_{\rm g} $, the coherence is also affected by the longitudinal interaction $ \Delta $, which is shown in Fig.~\ref{fig:3}. In Fig.~\ref{fig:3}, we see that the  results are similar to those of the excitation number discussed earlier.  Furthermore, the coherence $ \left<a\right> $  is also an order parameter of the QPT, which  can be viewed as the result of the $U(1)$ symmetry
	breaking~\cite{PhysRevLett.117.123602}. 
	\section{\label{sec:level5}PRACTICAL APPLICATION IN QUANTUM METROLOGY }  
	Now we consider the application in quantum metrology with the XXZ central spin model. Quantum criticality which can be viewed as a quantum resource has been widely applied to quantum sensing\cite{PhysRevA.78.042106,PhysRevA.78.042105,PhysRevLett.126.010502,PhysRevLett.124.120504,PhysRevLett.120.150501,PhysRevLett.96.140604,Ma2017}, and quantum Fisher information (QFI) is a key concept in quantum metrology, which gives the lower bound for the variance of the parameter estimation. In addition, the QFI proportional to fidelity susceptibility is used to quantify the abrupt change of the ground state in the vicinity of a critical point, and can be viewed as a good indicator for quantum phase transitions~\cite{PhysRevE.76.022101,doi:10.1142/S0217979210056633,PhysRevB.81.064418,PhysRevA.97.013845}.
	For a pure state $ \left|\psi\left(\tilde{g}\right)\right\rangle$ with a parameter $ \tilde{g} $, the  fidelity is $ f(\tilde{g},\delta{\tilde{g}})=\left|\left\langle \psi\left(\tilde{g}\right)|\psi\left(\tilde{g}+\delta \tilde{g}\right)\right\rangle \right| $ and the QFI is given by
	\begin{equation}\label{E17}
	F_{q}=-4\frac{\partial^2f(\tilde{g},\delta{\tilde{g}})}{\partial{(\delta \tilde{g}})^2}\Bigg|_{\delta \tilde{g}\rightarrow 0},
	\end{equation}
	where $ \delta{g}  $ is a small pertubation.
	
\begin{figure}
	
	\includegraphics[width=84mm,height=120mm]{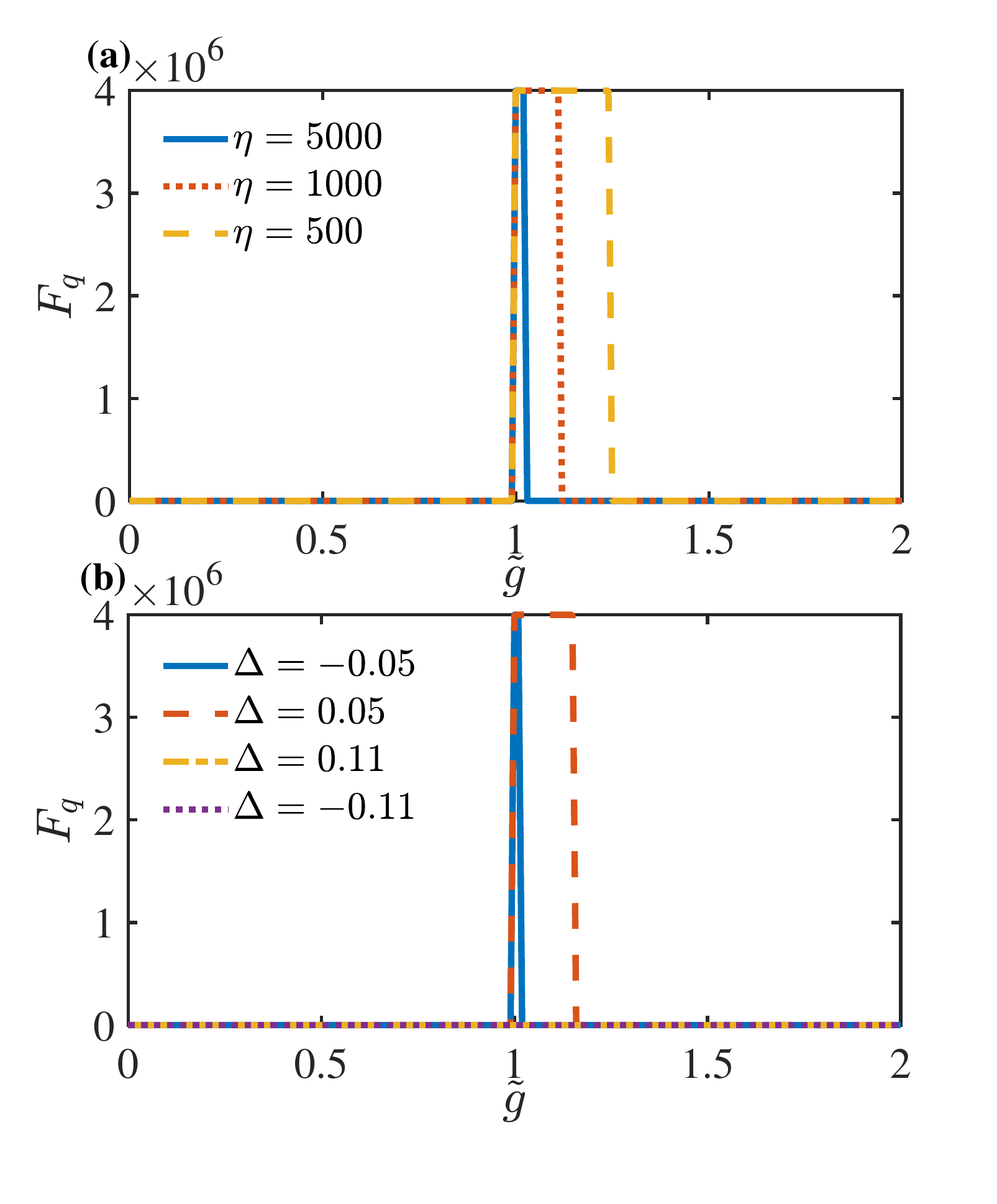}

	\caption{(a) QFI for different frequency ratios  $\eta$.  The width of the QFI becomes narrower as the frequency ratios increase. (b) QFI for different values of the longitudinal interaction $\Delta$. Here we set $ \omega=0.1 $ and $ \eta=10^{4} $. There exist abrupt changes of the QFI for $ \left|\Delta\right|<\omega $, however, for $ \left|\Delta\right|>\omega $, these abrupt changes disappear. }
	\label{fig:4}
\end{figure}

	In Fig.~\ref{fig:4}(a), we show the influence of different frequency ratios $\eta$ on the QFI. It is clear to see that the QFI has an abrupt change at the critical point $ \tilde{g}_{c}=1 $, which is due to a significant change in the ground state of the system. As the frequency ratio increases, the width of the QFI becomes narrower. Furthermore,  we use the fidelity approach to  verify the value range of the longitudinal interaction $\Delta$. In Fig.~\ref{fig:4} (b), we choose $ \omega=0.1$ and find that the abrupt changes of the QFI disappear when $\left|\Delta\right|>\omega $.
	
	In practical measurements, one can obtain the parameter information via suitable observables and the error propagation formula. In this model, the observable of the central spin $ \left\langle\sigma_{x}\right\rangle $ is able to measured, which is given by    
  	\begin{eqnarray}
  	\left\langle\sigma_{x}(t)\right\rangle&=&\left\langle\psi_{\rm in}\big|e^{iHt}\sigma_{x} e^{-iHt}\big|\psi_{\rm in}\right\rangle\nonumber\\
  	&=&2\mathrm{Re}
  	\left\{b_{\uparrow}^{*}b_{\downarrow}\left\langle \varphi\right|e^{it\left[2\widetilde{\omega}\left(\widetilde{g}^{2}+\frac{\Delta}{\widetilde{\omega}}\right)n\right]}\left|\varphi\right\rangle \right\}\label{E18}
  	\end{eqnarray}
	where $\left|\psi_{{\rm in}}\right\rangle =e^{s}\left(b_{\uparrow}\left|\uparrow\right\rangle +b_{\downarrow}\left|\downarrow\right\rangle \right)\otimes\left|\varphi\right\rangle  $ and $ \left|\varphi\right\rangle=\sum_{n}d_{n}\left|n\right\rangle$ with $\sum_{n}\left|d_{n}\right|^2=1  $.
	The inverse variance of the parameter $ \tilde{g} $ is given by \cite{PhysRevLett.126.010502}
	\begin{equation}\label{E19}
	\mathcal{F}_{\tilde{g}}=\frac{\left(\partial_{\tilde{g}}\left\langle \sigma_{x}\right\rangle \right)^{2}}{1-\left\langle \sigma_{x}\right\rangle ^{2}}.
	\end{equation}
	
	\begin{figure}
		
		\includegraphics[width=86mm,height=117mm]{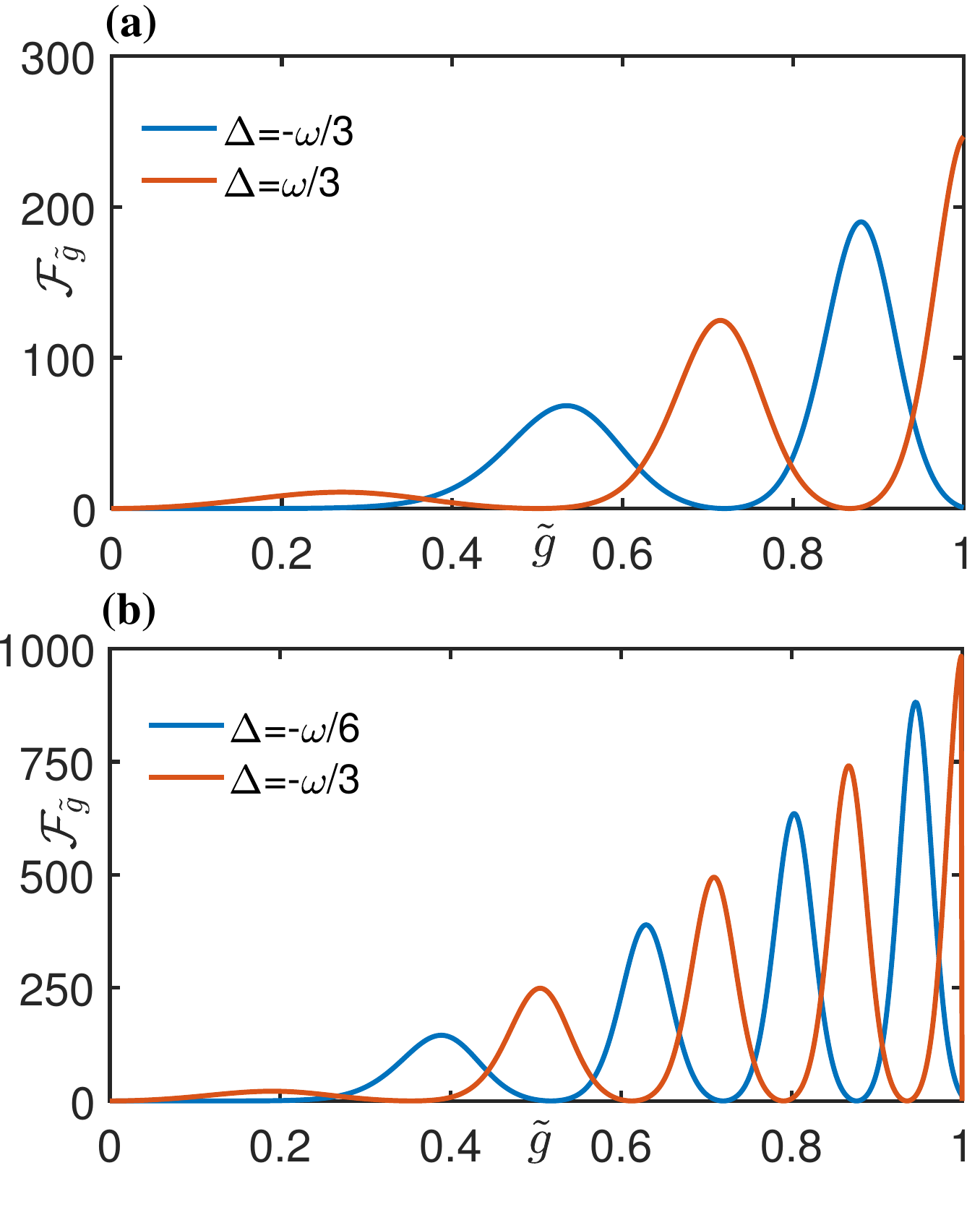}
		
		\caption{Inverse variance $ \mathcal{F}_{\tilde{g}} $ varies with $ \tilde{g} $. (a)  The evolution time is chosen as $ \tau=2\pi/\widetilde{\omega} $, and  different values of $\Delta$ change the peaks of $ \mathcal{F}_{\tilde{g}} $. As the coupling strength approaches the critical point, the amplitude of oscillation increases. (b) For $\tau=4\pi/\widetilde{\omega}$, the number of the peaks are twice that of the previous one. }
		\label{fig:5}
	\end{figure}
 In Fig.~\ref{fig:5}, we show the inverse variance $\mathcal{F}_{\tilde{g}}$ varies with $ \tilde{g} $. The initial state we choose is $b_{\uparrow}=b_{\downarrow}=1/\sqrt{2}$ and $  \varphi=\left|\alpha\right\rangle$. In spin systems, the spin coherent state can achieve the same effect with a large $N$~\cite{PhysRevA.87.052323,PhysRevB.99.174308}. It can be seen that $\mathcal{F}_{\tilde{g}}$ changes periodically, and the amplitude of oscillation increases as the coupling strength $\tilde{g}$ approaches the critical point. 
 It is worth noting that choosing different longitudinal interaction $\Delta$ has a significant influence on the positions of the maximum value and minimum values of $ \mathcal{F}_{g} $. It means we can change the coupling strength $ \tilde{g} $ at which maximum precision is achieved by controlling the value of the longitudinal interaction $\Delta$, instead of being confined to the vicinity of the critical point.

	\section{\label{sec:level6}CONCLUSION}
	In this paper, we have investigated the quantum phase transition in the XXZ central spin model, and the exact energy spectrum can be given analytically due to the $U(1)$ symmetry. In addition, the similarity between the JC model and  the central spin model is presented, and we have also demonstrated that the central spin model undergoes a superradiance QPT in the limit $ \eta\rightarrow\infty $ and $ N\rightarrow\infty $. To further explain the QPT in this model, we utilize the mean-field approximation to obtain the mean-field energy and the excitation number, which agree well with the numerical simulation.
	
	Moreover, the low-energy effective Hamiltonian given by the Schrieffer-Wolff transformation provides us with a new perspective to explain the quatum phase transition. Unlike the JC model, the nonlinear term caused by the longitudinal interaction can greatly affect the excitation number and the coherence of the ground state. For the case of $-\omega<\Delta<0$,  these two physical quantities increases significantly, while for the case of $ 0<\Delta<\omega $ they decreases remarkably. Furthermore, the QFI is used to  quantify the abrupt change of the ground state around the critical point, and we also utilize the error propagation formula to calculate the precision of the coupling strength. This work reveals the superradiant QPT occurring in central spin systems and provides a new idea for the realization of criticality-enhanced quantum sensing.

	\begin{acknowledgments}
		We acknowledge Dr.~Chao Han for valuable suggestions. This work was supported by the NSFC through Grant No.~11935012 and No.~11875231, the National Key Research and Development Program of China (Grants No.~2017YFA0304202 and No.~2017YFA0205700).
	\end{acknowledgments}
	\appendix
	
	\section{DERIVATION OF EXACT ENERGY SPECTRUM}\label{Appendix:A}
	
	Similar to the JC model, the dynamics of the XXZ central spin model is confined to the two-dimensional space spanned by  $\left|\uparrow,n-1\right>$ and $\left|\downarrow,n\right>$~\cite{gerry_knight_2004}. For a given $ n $, the matrix elements of $ H $ are 
	\begin{eqnarray}\nonumber
	\left\langle \uparrow,n-1\left|H\right|\uparrow,n-1\right\rangle &=&\frac{\omega_{0}}{2}+\left(\omega+\Delta\right)m,\\\nonumber
	\left\langle \downarrow,n\left|H\right|\downarrow,n\right\rangle &=&-\frac{\omega_{0}}{2}+\left(\omega-\Delta\right)\left(m+1\right),\\\nonumber
	\left\langle \uparrow,n-1\left|H\right|\downarrow,n\right\rangle &=&\left\langle \downarrow,n-1\left|H\right|\uparrow,n-1\right\rangle \\&=&A\sqrt{k_{n}},\label{A1}
	\end{eqnarray} 
	where $ m=\left(n-1-j\right) $ and $ k_{n}=\left(2j-n+1\right)n $.  The matrix representation is 
	\begin{equation}\label{A2}
	H=
	\begin{pmatrix}
	\frac{\omega_{0}}{2}+\left(\omega+\Delta\right)m&A\sqrt{k_{n}}\\
	A\sqrt{k_{n}}&-\frac{\omega_{0}}{2}+\left(\omega-\Delta\right)\left(m+1\right)
	\end{pmatrix}.
	\end{equation}
	We set $\Omega_{1}=2A\sqrt{k_{n}}$, $\Omega_{2}=(2m+1)\Delta-\omega+\omega_{0}$, $\Omega_{3}=\Delta-(2m+1)\omega$. Then we can obtain the energy eigenvalues given by
	\begin{equation}\label{A3}
	E_{\pm}=\frac{1}{2}\left(-\Omega_{3}\pm\sqrt{\Omega_{1}^2+\Omega_{2}^{2}}\right),
	\end{equation}
	and the eigenstates given by
	\begin{equation}\label{A4}
	\begin{split}
	\left|\psi_{+}\left(n\right)\right\rangle &=\widetilde{P}_{\uparrow,n-1}^{+}\left|\uparrow,n-1\right\rangle +\widetilde{P}_{\downarrow,n}^{+}\left|\downarrow,n\right\rangle, \\\left|\psi_{-}\left(n\right)\right\rangle &=\widetilde{P}_{\uparrow,n-1}^{-}\left|\uparrow,n-1\right\rangle +\widetilde{P}_{\downarrow,n}^{-}\left|\downarrow,n\right\rangle, 
	\end{split}
	\end{equation}
	where
	\begin{equation}\label{A5}
	\begin{split}
	\widetilde{P}_{\uparrow,n-1}^{\pm}&=\frac{\widetilde{\Omega}\pm\sqrt{1+\widetilde{\Omega}^{2}}}{\sqrt{2\left(1+\widetilde{\Omega}^{2}\right)\pm2\widetilde{\Omega}\sqrt{1+\widetilde{\Omega}^{2}}}},\\
	\widetilde{P}_{\downarrow,n\,\,\,}^{\pm}&=\frac{1}{\sqrt{2\left(1+\widetilde{\Omega}^{2}\right)\pm 2\Omega\sqrt{1+\widetilde{\Omega}^{2}}}},
	\end{split}
	\end{equation}
	and $\widetilde{\Omega}=\Omega_{2}/\Omega_{1}$.
	Now We use the mean-field approximation to get the mean-field energy. First, we apply the Holstein-Primakoff transformation to the Hamiltonian in Eq.~(\ref{E2}), where the angular momentum operators are represented by
	\begin{equation}\label{A6}
	\begin{split}
	J_{+}=\sqrt{N}a^{\dagger}\sqrt{1-\frac{a^{\dagger}a}{N}}&, \quad J_{-}=\sqrt{N}\sqrt{1-\frac{a^{\dagger}a}{N}},
	\\
	J_{z}=&a^{\dagger}a-\frac{N}{2}.
	\end{split}
	\end{equation}
	For large $N$, the Hamiltonian becomes 
	\begin{equation}\label{A7}
	H_{\rm hp}=\omega_{0}S_{z}+\omega \left(a^{\dagger}a-j\right)+\lambda\left(a^{\dagger}S_{-}+aS_{+}\right)+2\Delta\left(a^{\dagger}a-\frac{N}{2}\right)S_{z},
	\end{equation}
	where $\lambda=A\sqrt{2j}=A\sqrt{N}$. And the effective Hamiltonian under the mean-field approximation is given by
	\begin{equation}\label{A8}
	\begin{split}
	H_{\rm eff}&=\left\langle \beta|H_{\rm hp}|\beta\right\rangle\\
	&=\left(\widetilde{\omega}_{0}+2\Delta\left|\beta\right|^{2}\right)S_{z}+\omega\left(\left|\beta\right|^{2}-j\right)+\lambda\left(\beta^{*}S_{-}+\beta S_{+}\right),
	\end{split} 
	\end{equation}
	and the mean-field energy is
	\begin{equation}\label{A9}
	E_{-}=\omega\left(\left|\beta\right|^{2}-j\right)-\frac{1}{2}\overline{\omega}_{0},
	\end{equation}
	where $ \overline{\omega}_{0}(\left|\beta\right|^2)=\sqrt{4\lambda^{2}\left|\beta\right|^{2}+4\left|\beta\right|^{4}\Delta^{2}+4\left|\beta\right|^{2}\Delta\widetilde{\omega}_{0}+\widetilde{\omega}_{0}^{2}} $ is a function of $ \left|\beta\right|^2 $ and $ \widetilde{\omega}_{0}=\omega_{0}-N\Delta $. Here we define $ \tilde{g}=\lambda/\sqrt{\widetilde{\omega}\widetilde{\omega}_{0}} $ and  replace $\left|\beta\right|^{2}$ with $n$. Utilizing $\partial E_{-}/\partial n=0$, we find that for $ \tilde{g}<1 $ the excitation number of the ground state is still $n_{\rm{g}}=0$ and $E_{g}=-\widetilde{\omega}_{0}/2-\omega j$. However, for $ \tilde{g}>1$, we have 
	\begin{equation}\label{A10}
	n_{\rm{g}}=-\frac{\lambda^{2}+\text{\ensuremath{\Delta\widetilde{\omega}_{0}}}}{2\Delta^{2}}+\frac{\lambda\omega}{2\Delta^{2}}\sqrt{\frac{\lambda^{2}+2\Delta\widetilde{\omega}_{0}}{\omega^{2}-\Delta^{2}}}.
	\end{equation}
	and the energy of the  ground state under the mean-field approximation is 
	\begin{equation}\label{A11}
	E_{g}^{\mathrm {MF}}=\left(n_{\mathrm g}-j\right)\omega-\frac{1}{2}\overline{\omega}_{0}(n_{\rm g}),
	\end{equation}
	\setlength{\parskip}{0.5em}
	
	\section{DERIVATION OF LOW-ENERGY EFFECTIVE HAMILTONIAN}\label{Appendix:B}
	In this section, we give the derivation of the low-energy effective Hamiltonian in Eq.~(\ref{E9}). We first consider the case of the normal phase. $ H_{\rm hp} $ can be written as $ H_{\rm hp}=H_{0}+V$ where
	\begin{equation}\label{B1}
	\begin{split}
	H_{0}&=\widetilde{\omega}_{0}S_{z}+\omega \left(a^{\dagger}a-j\right)+2\Delta a^{\dagger}aS_{z},\\
	V&=\lambda\left(a^{\dagger}S_{-}+aS_{+}\right).
	\end{split}
	\end{equation} 
	 Now we use the method proposed in Refs.~\cite{PhysRevLett.115.180404,PhysRevLett.117.123602}. First we apply a Schrieffer-Wolff transformation $ e^{S} $ to $ H_{\rm hp} $, and the generator $ S $ is anti-Hermitian and block-off-diagonal. Then the Hamiltonian becomes 
	\begin{equation}\label{B2}
	\widetilde{H}=e^{-S}H_{\rm hp}e^{S}=\sum_{n=0}^{\infty}\frac{1}{n!}[H_{\rm hp},S]^{(n)},
	\end{equation}
	where $[H,S]^{(n)}=[[H,S]^{(n-1)},S]$ and $[H,S]^{(0)}=H$. Here we need the
	block-off-diagonal part of $ \widetilde{H} $ to be zero up to the second order in $ \lambda $, thus $ S $ must satisfies that
	\begin{equation}\label{B3}
	[H_{0},S]=-\lambda\left(a^{\dagger}S_{-}+aS_{+}\right).
	\end{equation}
	In the limit $\eta\rightarrow\infty$, we find that $S=\frac{\lambda}{\widetilde{\omega}_{0}}\left(a^{\dagger}S_{-}-aS_{+}\right) $, it leads to 
	\begin{equation}\label{B4}
	\begin{split}
	\widetilde{H}&=H_{0}+\frac{1}{2}\left[V,S\right]
	\\&=H_{0}+\frac{1}{2}\left[\lambda\left(a^{\dagger}S_{-}+aS_{+}\right),\frac{\lambda}{\widetilde{\omega}_{0}}\left(a^{\dagger}S_{-}-aS_{+}\right)\right]\\&=H_{0}+\frac{\lambda^{2}}{2\widetilde{\omega}_{0}}\left(4S_{z}a^{\dagger}a+2S_{z}+\sigma_{0}\right),
	\end{split}
	\end{equation}
	and the low-energy effective Hamiltonian is expressed as 
	\begin{equation}\label{B5}
	\begin{split}
	\widetilde{H}_{\rm np}&=\left\langle \downarrow\left|\widetilde{H}\right|\downarrow\right\rangle\\ 
	&=-\frac{\widetilde{\omega}_{0}}{2}-\omega j+\widetilde{\omega}\left(1-\tilde{g}^{2}\right)a^{\dagger}a,
	\end{split}
	\end{equation}
	where $\widetilde{\omega}=\omega-\Delta$, $\tilde{g}=\lambda/\sqrt{\widetilde{\omega}\widetilde{\omega}_{0}}  $.
	
	For the superradiant phase, we need to apply a  displacement operator $ D(\alpha) $ to $H_{\rm hp} $, which is given by
	\begin{eqnarray}\nonumber
    \overline{H}&=&D^{\dagger}\left(\alpha\right)H_{\rm hp}D\left(\alpha\right)\\\nonumber
	&=&\frac{\widetilde{\omega}_{0}}{2}\sigma_{z}+\omega a^{\dagger}a+\omega\alpha\left(a+a^{\dagger}\right)+\omega\alpha^{2}+\lambda\left(a^{\dagger}\sigma_{-}+a\sigma_{+}\right)\\
	&&+\lambda\alpha\sigma_{x}+\Delta\left(a^{\dagger}a+\alpha\left(a+a^{\dagger}\right)+\alpha^{2}\right)\sigma_{z},\label{B6}
    \end{eqnarray}
	where $D\left(\alpha\right)=e^{\alpha\left(a^{\dagger}-a\right)}$, $ \alpha^{2}=n_{\rm{g}}$, and here we make $ \alpha $ to be real for convenience. Now we get rid of the superscript of the Pauli operator and denote $ \sigma_{i}^{\left(0\right)}\equiv\sigma_{i}$ ($i=x,y,z$) for convenience. 
	
	We find that the part of the central spin in Eq.~(\ref{B6}) is  $ \left(\widetilde{\omega}_{0}+2\Delta\alpha ^{2}\right)\sigma_{z}/2+\lambda\alpha \sigma_{x} $, and its eigenstates are
	\begin{equation}\label{B7}
	\left|\tilde{\uparrow}\right\rangle =\cos\theta\left|\uparrow\right\rangle +\sin\theta\left|\downarrow\right\rangle, \left|\tilde{\downarrow}\right\rangle =-\sin\theta\left|\uparrow\right\rangle +\cos\theta\left|\downarrow\right\rangle, 
	\end{equation}
	where $ \theta=\frac{1}{2}\arctan\left(\frac{2\alpha\lambda}{2\alpha^{2}\Delta+\widetilde{\omega}_{0}}\right)$ . The corresponding eigenvalues are $ \pm\frac{\overline{\omega}_{0}(\alpha^2)}{2}=\pm\frac{1}{2}\sqrt{4\lambda^{2}\alpha ^{2}+4\alpha ^{4}\Delta^{2}+4\alpha ^{2}\Delta\widetilde{\omega}_{0}+\widetilde{\omega}_{0}^{2}} $.
	Note that we have $ \alpha^2=n_{\rm g} $, thus utilizing Eq.~(\ref{E8}) we have
	\begin{equation}\label{B8}
	\overline{\omega}_{0}(n_{\rm g})=\sqrt{\frac{\lambda^{4}+2\Delta\lambda^{2}\widetilde{\omega}_{0}}{\omega^{2}-\Delta^{2}}}.
	\end{equation}
	Then we use the eigenstates $ \left|\tilde{\uparrow}\right\rangle$($ \left|\tilde{\downarrow}\right\rangle  $) to rewrite Eq.~(\ref{B6}).
	\begin{eqnarray}\nonumber
	\overline{H}&=&
	\left(\frac{\lambda\widetilde{\omega}_{0}}{2\overline{\omega}_{0}}x-\frac{\lambda\Delta\alpha ^{2}}{\overline{\omega}_{0}}x-\frac{2\Delta\alpha \lambda}{\overline{\omega}_{0}}a^{\dagger}a\right)\tau_{x}\\\nonumber
	&&+\left(\frac{\overline{\omega}_{0}}{2}+\frac{\lambda^{2}\alpha}{\overline{\omega}_{0}}x+\frac{2\alpha^{3}\Delta^{2}+\Delta\alpha\widetilde{\omega}_{0}}{\overline{\omega}_{0}}x+\frac{2\alpha^{2}\Delta^{2}+\Delta\widetilde{\omega}_{0}}{\overline{\omega}_{0}}a^{\dagger}a\right)\tau_{z}\\
	&&-\frac{\lambda}{2}p\tau_{y}+\omega\alpha^{2}+\omega a^{\dagger}a+\omega\alpha x,\label{B9}
   \end{eqnarray}
	where $x=a^{\dagger}+a$ and $p=i\left(a^{\dagger}-a\right)$.
	 
	The Hamiltonian in Eq.~(\ref{B9}) can be divided into diagonal part $ \overline{H}_{0} $ and off-diagonal part $\overline{V}$, where 
	\begin{eqnarray}\nonumber
	\overline{H}_{0}&=&\omega a^{\dagger}a+\omega\alpha x+\omega\alpha ^{2}-\omega j\\\nonumber
	&&+\left(\frac{\overline{\omega}_{0}}{2}+\frac{\lambda^{2}\alpha }{\overline{\omega}_{0}}x+\frac{2\alpha ^{3}\Delta^{2}+\Delta\alpha \widetilde{\omega}_{0}}{\overline{\omega}_{0}}x+\frac{2\alpha ^{2}\Delta^{2}+\Delta\widetilde{\omega}_{0}}{\overline{\omega}_{0}}a^{\dagger}a\right)\tau_{z},\\\nonumber
	\overline{V}
	&=&\left(\frac{\lambda\widetilde{\omega}_{0}}{2\overline{\omega}_{0}}x-\frac{\lambda\Delta\alpha ^{2}}{\overline{\omega}_{0}}x-\frac{2\Delta\alpha \lambda}{\overline{\omega}_{0}}a^{\dagger}a\right)\tau_{x}-\frac{\lambda}{2}p\tau_{y}.\nonumber
	\end{eqnarray}
	\setlength{\parskip}{0.1em}
	Then we need to find the generator $ \overline{S} $ satisfies that $ [\overline{H}_{0},\overline{S}]=-\widetilde{V} $. In the limit $ \eta\rightarrow\infty $, $ \overline{S} $ is given by 
	\begin{equation}\label{B10}
	\overline{S}=\left(\frac{\lambda\widetilde{\omega}_{0}}{2i\overline{\omega}_{0}^{2}}x-\frac{\lambda\Delta\alpha ^{2}}{i\overline{\omega}_{0}^{2}}x-\frac{2\Delta\alpha \lambda}{i\overline{\omega}_{0}^{2}}a^{\dagger}a\right)\tau_{y}+\frac{\lambda}{2i\overline{\omega}_{0}}p\tau_{x}.
	\end{equation}
	The transformed Hamiltonian is
	\begin{widetext}
		\begin{eqnarray}\nonumber
		\overline{H}'&=&\overline{H}_{0}+\frac{1}{2}[\overline{V},\overline{S}]\\\nonumber
		&=&\omega a^{\dagger}a+\omega\alpha x+\omega\alpha ^{2}-\omega j+\left(\frac{\overline{\omega}_{0}}{2}+\frac{\lambda^{2}\alpha }{\overline{\omega}_{0}}x+\frac{2\alpha ^{3}\Delta^{2}+\Delta\alpha \widetilde{\omega}_{0}}{\overline{\omega}_{0}}x+\frac{2\alpha ^{2}\Delta^{2}+\Delta\widetilde{\omega}_{0}}{\overline{\omega}_{0}}a^{\dagger}a\right)\tau_{z}\\\nonumber
		&&+\left(\frac{\lambda^{2}\Delta^{2}\alpha ^{4}}{\overline{\omega}_{0}^{3}}-\frac{\lambda^{2}\widetilde{\omega}_{0}\Delta\alpha ^{2}}{\overline{\omega}_{0}^{3}}+\frac{\lambda^{2}\widetilde{\omega}_{0}^{2}}{4\overline{\omega}_{0}^{3}}\right)x^{2}\tau_{z}+\left(\frac{2\lambda^{2}\Delta^{2}\alpha ^{3}}{\overline{\omega}_{0}^{3}}-\frac{\lambda^{2}\Delta\alpha \widetilde{\omega}_{0}}{\overline{\omega}_{0}^{3}}\right)\left(xa^{\dagger}a+a^{\dagger}ax\right)\tau_{z}\\
		&&+\frac{4\Delta^{2}\alpha ^{2}\lambda^{2}}{\overline{\omega}_{0}^{3}}\left(a^{\dagger}a\right)^{2}\tau_{z}-\frac{\lambda^{2}\Delta\alpha ^{2}}{\overline{\omega}_{0}^{2}}+\frac{\lambda^{2}\widetilde{\omega}_{0}}{2\overline{\omega}_{0}^{2}}-\frac{\lambda^{2}\Delta\alpha }{\overline{\omega}_{0}^{2}}x+\frac{\lambda^{2}}{4\overline{\omega}_{0}}p^{2}\tau_{z},\label{B11}
		\end{eqnarray}
	\end{widetext}
	However, Eq.~(\ref{B11}) is too complicated to discuss further, thus we neglect the the terms with  $\overline{\omega}_{0}^{-3}$ and  $\overline{\omega}_{0}^{-2}$, and finally we obtain the low-energy effective Hamiltonian 
	\begin{eqnarray}\nonumber
	\overline{H}_{\rm sp}'&=&\left\langle \tilde{\downarrow}\Big|H^{'}\Big|\tilde{\downarrow}\right\rangle \\
	&=&\omega\alpha^{2}-\frac{\overline{\omega}_{0}}{2}+\kappa_{0}x^{2}+\kappa_{1}x+\kappa_{2}p^{2}-\omega j,\label{B13}
	\end{eqnarray}
	where 
	\begin{eqnarray}
	\kappa_{0}&=&\frac{2\alpha^{2}\Delta^{2}+\Delta\widetilde{\omega}_{0}}{4\overline{\omega}_{0}}-\frac{\omega}{4},\\
	\kappa_{1}&=&\omega\alpha-\frac{\lambda^{2}\alpha+2\left|\alpha\right|^{3}\Delta^{2}+\Delta\left|\alpha\right|\widetilde{\omega}_{0}}{\overline{\omega}_{0}},\\
	\kappa_{2}&=&\frac{\omega}{4}-\frac{2\alpha^{2}\Delta^{2}+\Delta\widetilde{\omega}_{0}+\lambda^{2}}{4\overline{\omega}_{0}}.
	\end{eqnarray}

	Utilizing Eq.~(\ref{E8}) and Eq.~(\ref{B8}), we find the relationship that $ 2\Delta^2\alpha^2+\lambda^2+\Delta\widetilde{\omega}_{0}=\omega\overline{\omega}_{0} $. It is easy to verify $ \kappa_{1}=\kappa_{2}=0 $ with the above relationship. Finally, we get Eq.~(\ref{E12}) in main text.
	\vspace{9pt}
	\bibliography{mybib}
\end{document}